\documentclass[10pt,letterpaper]{article}

\pdfoutput=1

\usepackage{opex3}
\usepackage{color}
\usepackage{threeparttable}

\usepackage[]{graphicx}
\usepackage{amsmath}  
\usepackage{amsfonts} 
\usepackage{bm}
\usepackage{subfigure}
\usepackage{textcomp}

\usepackage[mathscr]{euscript}

\usepackage{cite}

\usepackage{bigstrut}
\usepackage{cellspace}
\newcommand{\no}{\noindent}
\newcommand{\non}{\nonumber}
\newcommand{\lb}[1]{\label{#1}}
\newcommand{\beq}[1]{\begin{equation}\label{#1}}
\newcommand{\eeq}{\end{equation}}
\newcommand{\bseq}[1]{\begin{subequations}\label{#1}}
\newcommand{\eseq}{\end{subequations}}
\newcommand{\er}{\eqref}
\newcommand{\plr}[1]{\left(#1\right)}

\newcommand{\bklr}[1]{\left[#1\right]}
\newcommand{\abs}[1]{\left|#1\right|}

\newcommand{\ket}[1]{\left|{#1}\right\rangle}

\newcommand{\eq}{\equiv}

\newcommand{\dg}{\dagger}

\DeclareMathAlphabet\mathbfcal{OMS}{cmsy}{b}{n}

\begin{document}

\title{Polarization-based control of spin-orbit vector modes of light in biphoton interference}

\author{C.C. Leary,$^*$ Maggie Lankford, and Deepika Sundarraman}

\address{Department of Physics, The College of Wooster, Wooster, OH USA, 44691}

\email{$^*$cleary@wooster.edu} %% email address is required



\begin{abstract} 
We report the experimental generation of a class of spin-orbit vector modes of light via an asymmetric Mach-Zehnder interferometer, obtained from an input beam prepared in a product state of its spin and orbital degrees of freedom.  These modes contain a spatially varying polarization structure which may be controllably propagated about the beam axis by varying the retardance between the vertical and horizontal polarization components of the light.  Additionally, their transverse spatial intensity distributions may be continuously manipulated by tuning the input polarization parameters.  In the case of an analogous biphoton input, we predict that this device will exhibit biphoton (Hong-Ou-Mandel) interference in conjunction with the aforementioned tunable mode transformations.
\end{abstract}

 % REPLACE WITH CORRECT OCIS CODES FOR YOUR ARTICLE, MINIMUM OF TWO; Avoid using the OCIS codes for “General” or “General science” whenever possible.
%For a complete list of OCIS codes, visit: http://www.opticsinfobase.org/submit/ocis/

\bibliography{Leary_Lankford_Sundarraman_16_BibTex}
\bibliographystyle{osajnl}			% like plain but references appear in order of citation

\section{Introduction}\label{sec:intro} 

\begin{figure}
   \begin{center}
   \begin{tabular}{cc}
   \includegraphics[width=0.425\columnwidth]{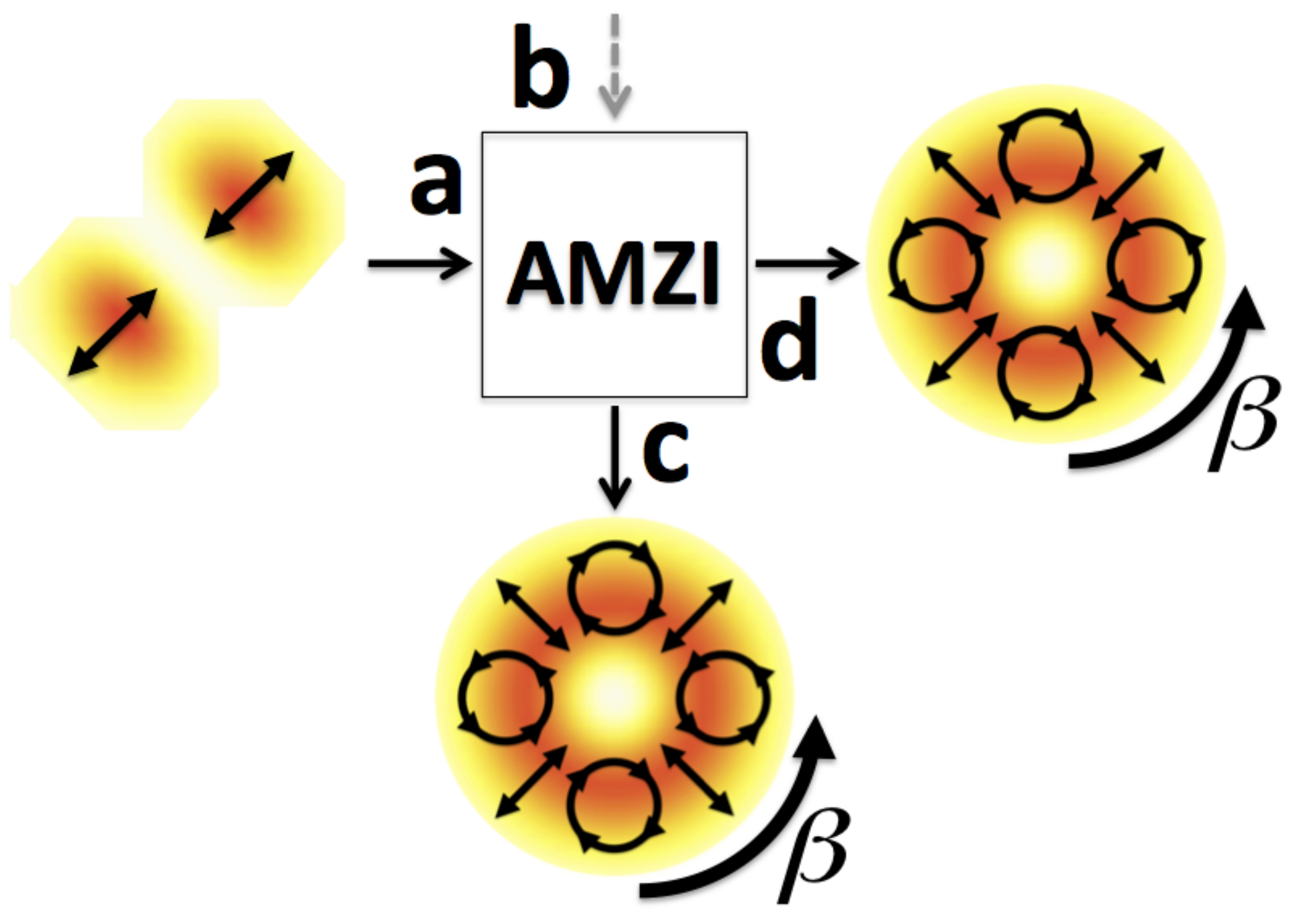} & \includegraphics[width=0.425\columnwidth]{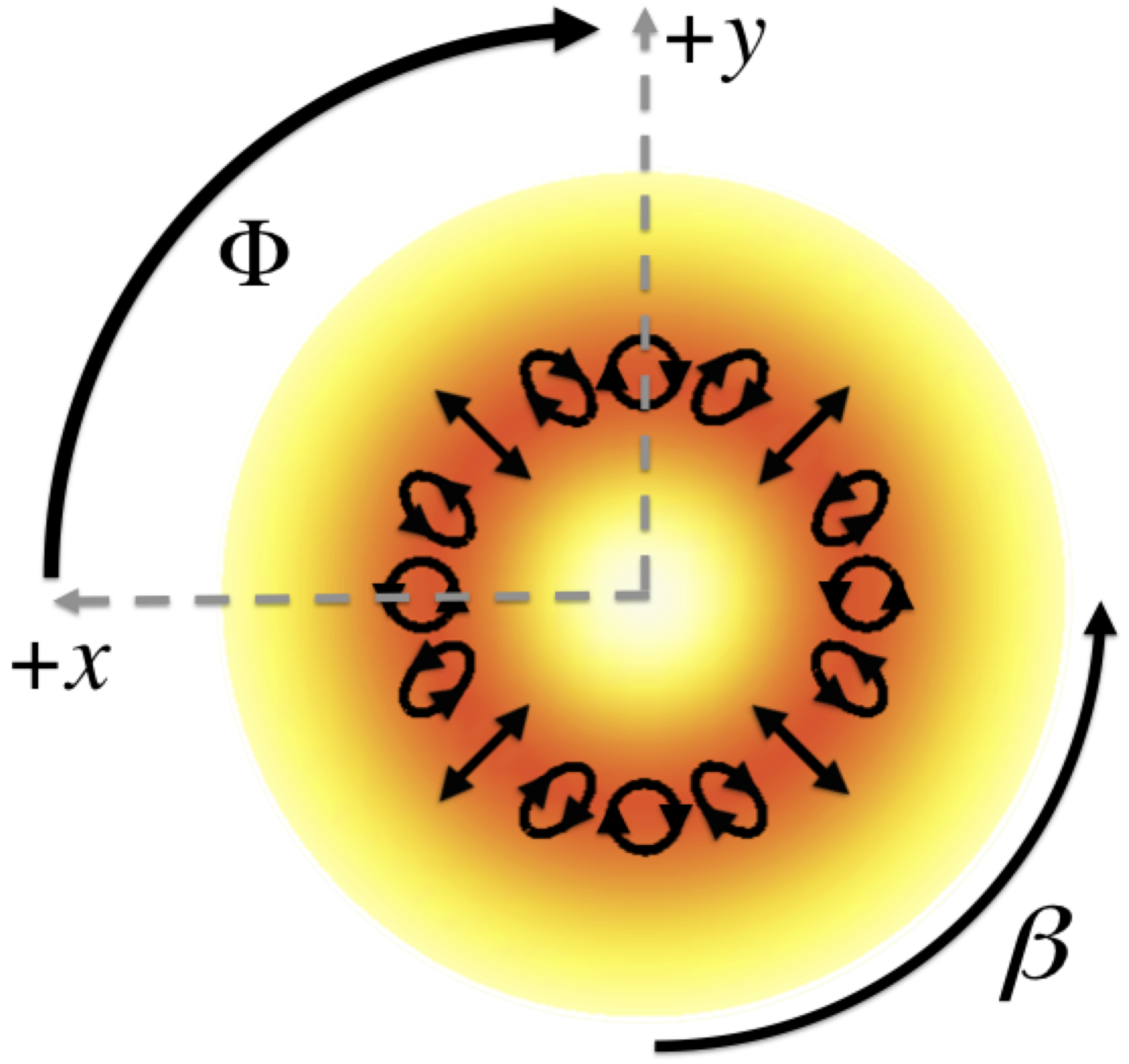} \\
$\;$ & $\;$ \\
   \text{\Large{(\textbf{a})}} &   \text{\Large{(\textbf{b})}}
   \end{tabular}
   \end{center}
   \vspace{-2.5mm}
   \caption 
   {\label{fig:Phenomenon} 
(a)  An input beam in a spin-orbit product state is converted to spin-orbit vector output modes by an asymmetric Mach-Zehnder interferometer (denoted as AMZI above).  
The shading is a density plot of the power density of each input and output mode, while the black lines represent the polarization states of these modes at various transverse spatial locations, as viewed from the beam source.  The polarization structure of each output mode propagates about the phase singularity as the retardance $\beta$ between vertical and horizontal polarization components is tuned.  (b)  Detail of the polarization structure of both output modes, whose polarization state exhibits variability in the azimuthal ($\Phi$) direction.  See Eq. \er{eq:BalancedOutput} and surrounding text for further details.}
\end{figure} 

The generation, measurement, and control of vector beams of light, whose transverse spatial and polarization degrees of freedom are nonseparable, have been the subject of a considerable amount of recent attention 
\cite{Maurer_2007,Souza_2007,Cardano_2012,Galvez_2012,Liu_2012,Gu_2013,Chen_2014,Yu_2015,Dambrosio_2015,Chen_2015,McLaren_2015,Khajavi_2015}.  Applications of vector beams in classical optics include polarimetry \cite{Tripathi_2009,Toppel_2014}, optical communication \cite{Wei_2015,Milione_2015b}, kinematic sensing \cite{Berg-Johansen_2015}, and optical trapping \cite{Zhang_2010,Skelton_2013,Zhou_2015,Milione_2015}.  Furthermore, the nonseparable nature of vector beams has allowed for experiments realizing local classical optics analogs to nonlocal quantum effects\cite{Aiello_2015}, including classical analogues to violations of Bell-like inequalities \cite{Borges_2010}, the Hardy test \cite{Karimi_2014}, and quantum teleportation \cite{Rafsanjani_2015}.  Within the purview of quantum optics and information, vector beams have been employed to demonstrate remote state preparation \cite{Barreiro_2010}, various forms of hybrid entanglement \cite{Karimi_2010,Fickler_2014}, spatially dependent electromagnetically induced transparency \cite{Radwell_2015}, and a multiple-degree-of-freedom quantum memory at the single-photon level \cite{Parigi_2015}.  In this context, the study of path-entangled biphoton states in which each photon's field mode is of the nonseparable type provides a natural direction of pursuit which has yet to be fully explored.

In this paper, we report the experimental generation of a class of nonseparable spin-orbit vector modes of light in an asymmetric Mach-Zehnder interferometer, and investigate theoretically the generation of path-entangled biphoton states with photons in nonseparable field modes within the same device.  These vector modes are obtained by passing laser light prepared in a product state of its spin and orbital degrees of freedom through an asymmetric Mach-Zehnder interferometer, with an extra mirror in one arm \cite{Sasada_2003}.  We show that these modes contain a spatially varying polarization structure which may be controllably propagated about the beam axis by varying the retardance between the vertical and horizontal polarization components of the light (see Fig. \ref{fig:Phenomenon} for a specific example of this phenomenon).  We additionally demonstrate the controlled manipulation of the transverse spatial intensity distribution of the interferometer output mode by tuning the polarization state of the input.  In the case of an analogous biphoton state in which spatially coherent light quanta from correlated photons pairs enter separate interferometer ports, we predict that this device will exhibit biphoton (Hong-Ou-Mandel) interference in conjunction with the aforementioned mode transformation:  that is, input photons in indistinguishable spin-orbit product states are converted to path-entangled output photons in nonseparable spin-orbit vector modes.  Although a number of works have studied biphoton interference involving the spin and/or orbital degrees of freedom (e.g. \cite{Walborn_2003,Yarnall_2007, Nagali_2009}), this phenomenon has, to our knowledge, yet to be realized for output photons sharing field modes whose internal degrees of freedom are nonseparable.  Potential applications of the tunable vector mode structures discussed in this work involve their use in spatially inhomogeneous light-matter interactions which depend on the polarization degree of freedom of light, including the design of dynamic optical tweezers on the classical beam level, and interactions with atomic systems involving biphoton states of light in structured vector modes.

This work is organized as follows: In Section \ref{sec:Prelims}, we parameterize the spin-orbit product input states of interest in terms of their associated spin and orbital Poincar\'e sphere angles \cite{Padgett_1999}, and describe the action of the asymmetric Mach-Zehnder interferometer as a unitary operation on these states.  In Section \ref{sec:Biphoton}, we predict the simultaneous biphoton interference and mode conversion effects mentioned above, giving conditions under which biphoton interference takes place along with expressions for the biphoton output state and associated nonseparable spin-orbit output field modes.  In Section \ref{sec:ModeFunctions}, we contrast this quantum optical treatment with the action of our interferometer in the corresponding classical picture, and provide a theoretical description of the predicted output mode functions in terms of the spin and orbital Poincar\'e sphere angles of the input beams.  Finally, in Section \ref{sec:Experiment}, we compare classical predictions to experimental data for several special cases of the theory, thereby demonstrating the aforementioned effects related to the polarization-based control of vector mode distributions of light.

\section{Preliminaries}\lb{sec:Prelims}

   \begin{figure}
   \begin{center}
   \includegraphics[width=1.0\columnwidth]{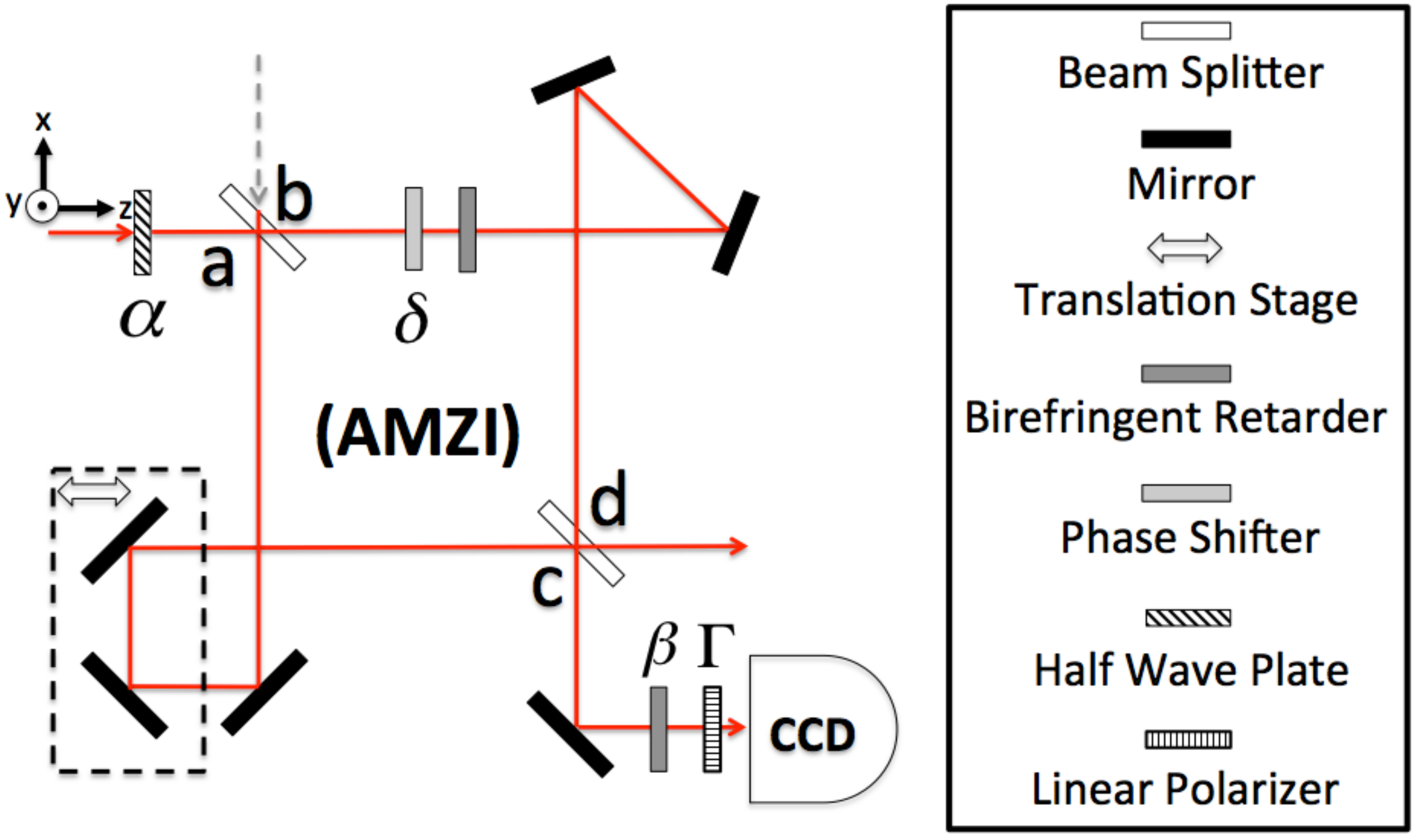} \\
   \end{center}
   \vspace{-3mm}
   \caption
   {\label{fig:Interferometer} A four-port asymmetric Mach-Zehnder interferometer with balanced path lengths and internal phase shifter $\delta$.  The linearly polarized input beam is prepared via a half wave plate, and the output passes through a birefringent retarder and/or a linear polarizer before measurement via CCD camera.  See the key above and main text for additional details.}  
   \end{figure}

\subsection{Spin-orbit product mode inputs}

Consider a four port asymmetric Mach-Zehnder interferometer with an extra mirror in one arm, which consists of two input ports $a$ and $b$ accepting mutually coherent classical electromagnetic fields $\mathbf{E}_a$ and $\mathbf{E}_b$ along with two output ports $c$ and $d$ transmitting fields $\mathbf{E}_c$ or $\mathbf{E}_d$, as shown in Fig. \ref{fig:Interferometer}.  We will assume throughout this work that the translation stage shown in the figure has been adjusted such that the interferometer has equal path lengths.  We further assume that the input fields (and therefore also the output fields) are collimated monochromatic Gaussian laser beams sharing a common frequency $\omega$, wave number $k$, and a constant effective beam radius $w_0$, and define ``local'' cartesian basis vectors $\hat{\mathbf{x}},\hat{\mathbf{y}},\hat{\mathbf{z}}$ as shown in the figure, with cylindrical coordinates $\plr{r,\Phi,z}$ attached to each beam axis.  Under these conditions, all input and output fields share the following general traveling wave form, $\mathbf{E}_p\plr{r,z,t}=\mathbfcal{E}_p\plr{r,\Phi}\exp\bklr{{i\plr{kz-\omega t}}}$, where $\mathbfcal{E}_p\plr{r,\Phi}$ contains the transverse spatial and vector dependence of the field, with the index $p$ taking one of the four values $a,b,c,$ or $d$ depending on the port of interest. 

We assume that $\mathbfcal{E}_a\plr{r,\Phi}$ and $\mathbfcal{E}_b\plr{r,\Phi}$ are in product states of their transverse spatial (orbital) and polarization (spin) degrees of freedom, so that a single input photon in such a field mode may encode two qubits, with one coded within its orbital degrees of freedom and the other within its spin.  In order to facilitate a comparison of the present classical description with the case of biphoton quantum interference, we further assume that the orbital and spin qubits associated with each input field are mutually indistinguishable.  Under the above conditions, we may express the input fields as

\beq{eq:Product}
\mathbfcal{E}_p\plr{r,\Phi}=E_{p}\,\mathcal{E}\plr{r,\Phi}\bm{\hat{e}}, \;\;\;\;\; \;\;\;\;\; \plr{p=a,b}.
\eeq

\no where the constant scalar $E_{p}$ denotes the field amplitude for a given port $p$.  For the orbital qubit in Eq. \er{eq:Product}, we constrain the transverse spatial distribution of each input field to take the form a linear superposition of first-order Hermite-Gaussian basis modes \cite{Padgett_1999}, so that 

\beq{eq:SpatialSup}
\mathcal{E}\!\plr{r,\Phi}=\cos{\!\plr{\tfrac{\theta}{2}}}\mathcal{E}_0+\sin{\!\plr{\tfrac{\theta}{2}}}e^{i \phi} \mathcal{E}_1,
\eeq

\no where $\mathcal{E}_0\eq G\plr{r}\sin{\Phi}$ and $\mathcal{E}_1\eq G\plr{r}\cos{\Phi}$ respectively denote the $HG_{01}$ and $HG_{10}$ modes, with the common Gaussian function $G\plr{r}\eq \sqrt{\tfrac{8}{\pi}}\tfrac{r}{w_0^2}\exp\bklr{{-\plr{\tfrac{r}{w_0}}^2}}$ normalized such that $\int_0^\infty\int_0^\pi\mathcal{E}_0^2rdrd\Phi=\int_0^\infty\int_0^\pi\mathcal{E}_1^2rdrd\Phi=1$.  We similarly encode the spin qubit of each input field in the form of a superposition of vertical and horizontal linear polarization states,

\beq{eq:PolSup}
\bm{\hat{e}}=\cos{\!\plr{\tfrac{\alpha}{2}}}\bm{\hat{e}}_0+\sin{\!\plr{\tfrac{\alpha}{2}}}e^{i \beta}\bm{\hat{e}}_1.
\eeq

\no where $\bm{\hat{e}}_0\eq\bm{\hat{y}}$ denotes the direction of vertical polarization and $\bm{\hat{e}}_1\eq\bm{\hat{x}}$ the horizontal.

In Eq. \er{eq:SpatialSup}, $\plr{\theta, \phi}$ are spherical coordinates parameterizing the first-order Hermite-Gaussian spatial mode Poincar\'e sphere, while in Eq. \er{eq:PolSup}, the angles $\plr{\alpha, \beta}$ parameterize the analogous polarization Poincar\'e sphere \cite{Padgett_1999}.  With these preliminaries, substitution of Eqs. \er{eq:PolSup} and \er{eq:SpatialSup} into Eq. \er{eq:Product} yields the following expression for the input modes,

\beq{eq:Inputs}
\mathbfcal{E}_p\plr{r,\Phi}=e_{00}^{p}\,\mathbfcal{E}_{00}\!+e_{11}^{p}\,\mathbfcal{E}_{11}\!+e_{10}^{p}\,\mathbfcal{E}_{10}\!+e_{01}^{p}\,\mathbfcal{E}_{01} ,
\eeq

\no where

\begin{align}\lb{eq:Amplitudes}
e_{00}^{p}&\eq E_p\cos{\!\plr{\tfrac{\theta}{2}}}\cos{\!\plr{\tfrac{\alpha}{2}}}, \;\;\;\; \;\;\;\; \;\;\;\; \;\;\;\; \;\;\;\; \;\;\;\; \;\;\; \mathbfcal{E}_{00}\!\eq\mathcal{E}_0\bm{\hat{e}}_0= G\plr{r}\sin{\!\Phi}\,\bm{\hat{y}}, \non \\
e_{11}^{p}&\eq E_p\sin{\!\plr{\tfrac{\theta}{2}}}\sin{\!\plr{\tfrac{\alpha}{2}}}e^{i\plr{\phi+\beta}}, \;\;\;\; \;\;\;\; \;\;\;\; \;\;\;\; \mathbfcal{E}_{11}\!\eq\mathcal{E}_1\bm{\hat{e}}_1= G\plr{r}\cos{\!\Phi}\,\bm{\hat{x}}, \non \\
e_{10}^{p}&\eq E_p\sin{\!\plr{\tfrac{\theta}{2}}}\cos{\!\plr{\tfrac{\alpha}{2}}}e^{i\phi}, \;\;\;\; \;\;\;\; \;\;\;\; \;\;\;\; \;\;\;\; \;\;\, \mathbfcal{E}_{10}\!\eq\mathcal{E}_1\bm{\hat{e}}_0= G\plr{r}\cos{\!\Phi}\,\bm{\hat{y}}, \non \\
e_{01}^{p}&\eq E_p\cos{\!\plr{\tfrac{\theta}{2}}}\sin{\!\plr{\tfrac{\alpha}{2}}}e^{i\beta}, \;\;\;\; \;\;\;\; \;\;\;\; \;\;\;\; \;\;\;\; \;\;\, \mathbfcal{E}_{01}\!\eq\mathcal{E}_0\bm{\hat{e}}_1= G\plr{r}\sin{\!\Phi}\,\bm{\hat{x}}.
\end{align}

\no Here, the $e_{jk}^{p}$ coefficients denote normalized amplitudes for the respective transverse spatial vector basis modes $\mathbfcal{E}_{jk}$ comprising the input field at port $p=a$ or $b$.  The $j$ and $k$ indices correspond respectively to the properties of the transverse spatial mode function $\mathcal{E}\plr{r,\Phi}$ and polarization vector $\bm{\hat{e}}$ under one-dimensional parity inversion of the $x$ axis.  A one-dimensional parity inversion of the $x$ axis involves both the coordinate transformation $x\to-x$ (which is equivalent to the transformation $\Phi\to-\Phi$ in cylindrical coordinates) and the vector field transformation $\bm{\hat{x}}\to-\bm{\hat{x}}$.  Therefore, $j=0/1$ denotes even/odd parity of $\mathcal{E}\plr{r,\Phi}$ upon x-inversion, and $k=0/1$ similarly denotes even/odd parity of $\bm{\hat{e}}$.  Since the vector modes $\mathbfcal{E}_{jk}$ given in Eq. \er{eq:Amplitudes} span the entire mode space of interest, we choose to employ this basis to represent the output fields as well, so that Eq. \er{eq:Inputs} describes the fields at all four ports.  However, the output amplitudes $e_{jk}^{c}$ and $e_{jk}^{d}$ do not generally possess the simple product form Eq. \er{eq:Amplitudes}, which applies only to the input amplitudes.

\subsection{Asymmetric Mach-Zehnder interferometer} 

Each time light reflects off a mirror or beam splitter in the interferometer in Fig. \ref{fig:Interferometer}, it picks up a polarization-dependent phase shift, such that the total phase shift accumulated by each polarization component of a given input wave depends on the interferometer arm traversed.  In order to compensate for these shifts in our experiment, we insert birefringent retarders in the upper interferometer arm as well as at the output (see Fig. \ref{fig:Interferometer}).  Additionally, we insert an internal phase shifter in the form of a thin glass plate into the upper arm as shown, in order to control the relative phase $\delta$ between interferometer arms independently of polarization.

With the polarization-related phase compensators appropriately set, the relationship between the input and output basis mode amplitudes $e_{jk}^{p}$ may be modeled up to an irrelevant overall constant phase by connecting them by means of a series of unitary transformations \cite{Sasada_2003}, respectively representing the actions of the first beam spitter, the internal phase shifter $\delta$, the mirrors, and the final beam splitter:

\arraycolsep=2pt

\begin{align} \lb{eq:U}
\left(
\begin{array}{c} {e_{jk}^{c}}  \\ {e_{jk}^{d}} \end{array}\right)
&=\frac{1}{\sqrt{2}}\left(\begin{array}{cc} {1} & {i\Pi_{jk}} \\ {i\Pi_{jk}} & {1} \end{array}\right)
\left(\begin{array}{cc} {\Pi_{jk}^2} & {0} \\ {0} & {\Pi_{jk}^3} \end{array}\right)
\left(\begin{array}{cc} {e^{i\frac{\delta}{2}}} & {0} \\ {0} & {e^{-i\frac{\delta}{2}}} \end{array}\right)
\frac{1}{\sqrt{2}}\left(\begin{array}{cc} {1} & {i\Pi_{jk}} \\ {i\Pi_{jk}} & {1} \end{array}\right) 
\left(\begin{array}{c} {e_{jk}^{a}}  \\ {e_{jk}^{b}} \end{array}\right), \non \\
&=\frac{1}{2}\left(\begin{array}{cc} {e^{i\frac{\delta}{2}}-\Pi_{jk}e^{-i\frac{\delta}{2}}} & {i\Pi_{jk}(e^{i\frac{\delta}{2}}+\Pi_{jk}e^{-i\frac{\delta}{2}})} \\ {i\Pi_{jk}(e^{i\frac{\delta}{2}}+\Pi_{jk}e^{-i\frac{\delta}{2}})} & {-(e^{i\frac{\delta}{2}}-\Pi_{jk}e^{-i\frac{\delta}{2}})} \end{array}\right)
\left(\begin{array}{c} {e_{jk}^{a}}  \\ {e_{jk}^{b}} \end{array}\right)
\eq\hat{U}_{jk}
\left(\begin{array}{c} {e_{jk}^a}  \\ {e_{jk}^b} \end{array}\right).
\end{align}

\no Here, $\Pi_{jk}\eq\plr{-1}^{j+k}$ represents a reflective phase shift that depends on the one-dimensional parity properties associated with the basis mode $\mathbfcal{E}_{jk}$ upon reflection about the $x$ axis.  This phase shift occurs at each physical reflection in our apparatus, from both mirrors and beam splitters. For transverse spatial vector basis modes that possess even parity upon reflection, $j+k$ is an even quantity, while for odd-parity modes $j+k$ is odd (cf. Eq. \er{eq:Amplitudes}).  Therefore, in Eq. \er{eq:U} we have

\begin{align}\lb{eq:EvenOdd}
&\hat{U}_{jk}=i\left(\begin{array}{cc} {\sin{\plr{\frac{\delta}{2}}}} & {\cos{\plr{\frac{\delta}{2}}}} \\ {\cos{\plr{\frac{\delta}{2}}}} & {-\sin{\plr{\frac{\delta}{2}}}} \end{array}\right) \;\;\;\;\; \;\;\;\;\;  \hat{U}_{jk}=\left(\begin{array}{cc} {\cos{\plr{\frac{\delta}{2}}}} & {\sin{\plr{\frac{\delta}{2}}}} \\ {\sin{\plr{\frac{\delta}{2}}}} & {-\cos{\plr{\frac{\delta}{2}}}} \end{array}\right) \;\;\;\;\; \;\;\;\;\; \\
&\text{for $j+k$ even}, \;\;\;\;\; \;\;\;\;\ \;\;\;\;\; \;\;\;\;\ \;\;\;\;\; \;\;\;\;\ \;\;\;\;\; \;\;\;\;\ \;\;\;\, \text{for $j+k$ odd}. \non
\end{align}

Equations \er{eq:U} and \er{eq:EvenOdd} provide the connection between the input and output spin-orbit field modes in both the classical and quantum optical descriptions of interference within our experimental device.  In what follows we develop the associated quantum description, which predicts biphoton interference to occur in conjunction with a mode conversion from product state input photons to spin-orbit vector output photons exhibiting nonseparability of spin and orbital degrees of freedom in their associated field mode functions.  We then proceed to analyze these mode functions both theoretically and experimentally within a classical framework.

\section{Biphoton quantum interference with conversion to vector modes}\lb{sec:Biphoton}

In a restricted ``four-mode'' approach to quantization, the transverse spatial vector field functions $\mathbfcal{E}_p\plr{r,\Phi}$ are quantized by promoting the mode function coefficients of Eq. \er{eq:U} to operators according to the prescription ${(e_{jk}^{p})}^*\to \mathcal{E}\hat{p}_{jk}^{\dg}$.  Here, the asterisk symbol denotes the complex conjugate, $\hat{p}_{jk}^{\dg}$ denotes the Fock-space creation operator for a photon at port $p=a,b,c$,~or~$d$ and in field mode $\mathbfcal{E}_{jk}$, while $\mathcal{E}$ is a quantization constant with units of electric field, the specific form of which is irrelevant to our purposes.  The creation operators $\hat{p}_{jk}^{\dg}$ obey the usual commutation relations with their associated annihilation operators.  If correlated biphoton pairs prepared in indistinguishable spatially coherent spin-orbit product modes of the form given in Eq. \er{eq:Inputs} are allowed to impinge on input ports $a$ and $b$ such that the photons overlap longitudinally to within one another's coherence lengths, then the resulting (unnormalized) biphoton input state $\ket{\psi_{\text{in}}}$ is given by

\begin{align}\lb{eq:Biphoton}
\ket{\psi_{\text{in}}}&=\plr{e_{00}\,\hat{a}_{00}^{\dg}+e_{11}\,\hat{a}_{11}^{\dg}+e_{10}\,\hat{a}_{10}^{\dg}+e_{01}\,\hat{a}_{01}^{\dg}} \!\!
\plr{e_{00}\,\hat{b}_{00}^{\dg}+e_{11}\,\hat{b}_{11}^{\dg}+e_{10}\,\hat{b}_{10}^{\dg}+e_{01}\,\hat{b}_{01}^{\dg}}\ket{\text{vac}}.
\end{align}

\no Here, each creation operator $\hat{p}_{jk}^{\dg}$ has been weighted according to the amplitude $e_{jk}$ of its associated mode function $\mathbfcal{E}_{jk}$ (cf. Eq. \er{eq:Inputs}), with the $e_{jk}$ coefficients given by Eq. \er{eq:Amplitudes}, but with the classical field amplitudes $E_{p}$ dropped.  Also in Eq. \er{eq:Biphoton}, $\ket{\text{vac}}$ denotes the vacuum state, and we have omitted the common quantization factor $\mathcal{E}$.  We are interested in the biphoton output state $\ket{\psi_{\text{out}}}$ resulting from the action of the asymmetric interferometer on the input state $\ket{\psi_{\text{in}}}$.  Toward this end, substituting Eqs. \er{eq:EvenOdd} into Eq. \er{eq:U}, inverting the results, and then conjugating both sides in order to carry out the aforementioned  quantization procedure ${(e_{jk}^{p})}^*\to \mathcal{E}\hat{p}_{jk}^{\dg}$ leads to the following input-output operator relations,

\arraycolsep=2pt
\bseq{eq:OperatorRelations}
\begin{align}
&\left(\begin{array}{c} {\hat{a}_{jk}^{\dg}}  \\ {\hat{b}_{jk}^{\dg}} \end{array}\right)=
i\left(\begin{array}{cc} {\sin{\plr{\frac{\delta}{2}}}} & {\cos{\plr{\frac{\delta}{2}}}} \\ {\cos{\plr{\frac{\delta}{2}}}} & {-\sin{\plr{\frac{\delta}{2}}}} \end{array}\right)
\left(\begin{array}{c} {\hat{c}_{jk}^{\dg}}  \\ {\hat{d}_{jk}^{\dg}}\end{array}\right) \;\;\;\;\;\;\;\; \left(\begin{array}{c} {\hat{a}_{jk}^{\dg}}  \\ {\hat{b}_{jk}^{\dg}} \end{array}\right)=
\left(\begin{array}{cc} {\cos{\plr{\frac{\delta}{2}}}} & {\sin{\plr{\frac{\delta}{2}}}} \\ {\sin{\plr{\frac{\delta}{2}}}} & {-\cos{\plr{\frac{\delta}{2}}}} \end{array}\right)
\left(\begin{array}{c} {\hat{c}_{jk}^{\dg}}  \\ {\hat{d}_{jk}^{\dg}}\end{array}\right)  \non \\
& \;\; \text{for $j+k$ even}, 
\;\;\;\; \;\;\;\; \;\;\;\; \;\;\;\; \;\;\;\; \;\;\;\; \;\;\;\; \;\;\;\; \;\;\;\; \;\;\;\; \;\;\;\; \;\;\;\; \;\;\;\; \;\;\;\; \;\;\;\;  \;\;\;  \text{for $j+k$ odd}.
\end{align}
\eseq

\no Substituting Eqs. \er{eq:OperatorRelations} into Eq. \er{eq:Biphoton} and grouping terms according to output port then yields

\bseq{eq:BiphotonOutput}
\begin{align}
\ket{\psi_{\text{out}}}&\propto\plr{\hat{\Phi}_c^{\dg}+\hat{\Phi}_d^{\dg}}
\!\!\plr{\hat{\mathscr{X}}_c^{\dg}-\hat{\mathscr{X}}_d^{\dg}}\ket{\text{vac}}=\plr{\hat{\Phi}_c^{\dg}\hat{\mathscr{X}}_c^{\dg}-\hat{\Phi}_d^{\dg}\hat{\mathscr{X}}_d^{\dg}
-\hat{\Phi}_c^{\dg}\hat{\mathscr{X}}_d^{\dg}+\hat{\Phi}_d^{\dg}\hat{\mathscr{X}}_c^{\dg}}\ket{\text{vac}}, \lb{eq:BiphotonOutputB}
\end{align}
\eseq

\no where

\bseq{eq:OperatorDefinitions}
\begin{align}
\hat{\Phi}_c^{\dg}&\!\eq\!\sin\!{\plr{\tfrac{\delta}{2}}}\!\!\plr{\!e_{00}\,\hat{c}_{00}^{\dg}\!+\!e_{11}\,\hat{c}_{11}^{\dg}\!}
\!-\!i\cos\!{\plr{\tfrac{\delta}{2}}}\!\!\plr{\!e_{10}\,\hat{c}_{10}^{\dg}\!+\!e_{01}\,\hat{c}_{01}^{\dg}\!}\!, \\
\hat{\mathscr{X}}_c^{\dg}&\!\eq\!\cos\!{\plr{\tfrac{\delta}{2}}}\!\!\plr{\!e_{00}\,\hat{c}_{00}^{\dg}\!+\!e_{11}\,\hat{c}_{11}^{\dg}\!}
\!-\!i\sin\!{\plr{\tfrac{\delta}{2}}}\!\!\plr{\!e_{10}\,\hat{c}_{10}^{\dg}\!+\!e_{01}\,\hat{c}_{01}^{\dg}\!}\!, \\
\hat{\Phi}_d^{\dg}&\!\eq\!\cos\!{\plr{\tfrac{\delta}{2}}}\!\!\plr{\!e_{00}\,\hat{d}_{00}^{\dg}\!+\!e_{11}\,\hat{d}_{11}^{\dg}\!}
\!-\!i\sin\!{\plr{\tfrac{\delta}{2}}}\!\!\plr{\!e_{10}\,\hat{d}_{10}^{\dg}\!+\!e_{01}\,\hat{d}_{01}^{\dg}\!}\!, \\
\hat{\mathscr{X}}_d^{\dg}&\!\eq\!\sin\!{\plr{\tfrac{\delta}{2}}}\!\!\plr{\!e_{00}\,\hat{d}_{00}^{\dg}\!+\!e_{11}\,\hat{d}_{11}^{\dg}\!}
\!-\!i\cos\!{\plr{\tfrac{\delta}{2}}}\!\!\plr{\!e_{10}\,\hat{d}_{10}^{\dg}\!+\!e_{01}\,\hat{d}_{01}^{\dg}\!}\!. 
\end{align}
\eseq

By inspection of Eqs. \er{eq:BiphotonOutput} and \er{eq:OperatorDefinitions}, it is evident that Hong-Ou-Mandel interference occurs when the internal interferometer phase $\delta$ is set to one of the two values $\delta=\pm\tfrac{\pi}{2}$.  Under these circumstances, $\hat{\Phi}_c^{\dg}=\hat{\mathscr{X}}_c^{\dg}$ and $\hat{\Phi}_d^{\dg}=\hat{\mathscr{X}}_d^{\dg}$, from which it follows that $\hat{\Phi}_c^{\dg}\hat{\mathscr{X}}_d^{\dg}=\hat{\Phi}_d^{\dg}\hat{\mathscr{X}}_c^{\dg}$ so that the latter two terms in Eq. \er{eq:BiphotonOutputB} cancel, yielding the biphoton output state

\begin{align}\lb{eq:BiphotonOutputState}
\ket{\psi_{\text{out}}}&\propto\plr{\hat{\Psi}_{c_{\mp}}^{{\dg}^2}-\hat{\Psi}_{d_{\mp}}^{{\dg}^2}}\ket{\text{vac}}=\ket{2_{c,\bm{\Psi}_{\mp}}0_{d,\bm{\Psi}_{\mp}}}-\ket{0_{c,\bm{\Psi}_{\mp}}2_{d,\bm{\Psi}_{\mp}}}. 
\end{align}

\no In Eq. \er{eq:BiphotonOutputState},

\bseq{eq:HOMIModes}
\begin{align}
\hat{\Psi}_{c_{\mp}}^{{\dg}^2}&\eq \bklr{\plr{e_{00}\,\hat{c}_{00}^{\dg}+e_{11}\,\hat{c}_{11}^{\dg}}\mp i\plr{e_{10}\,\hat{c}_{10}^{\dg}+e_{01}\,\hat{c}_{01}^{\dg}}}^2, \\
\hat{\Psi}_{d_{\mp}}^{{\dg}^2}&\eq \bklr{\plr{e_{00}\,\hat{d}_{00}^{\dg}+e_{11}\,\hat{d}_{11}^{\dg}}\mp i\plr{e_{10}\,\hat{d}_{10}^{\dg}+e_{01}\,\hat{d}_{01}^{\dg}}}^2, 
\end{align}
\eseq

\no while the Fock space ket vectors $\ket{2_{c,\bm{\Psi}_{\mp}}0_{d,\bm{\Psi}_{\mp}}}$ and $\ket{0_{c,\bm{\Psi}_{\mp}}2_{d,\bm{\Psi}_{\mp}}}$ denote two-photon Fock states associated with the creation operators given in Eq. \er{eq:HOMIModes}.  In both Eqs. \er{eq:BiphotonOutputState} and \er{eq:HOMIModes}, the upper and lower signs in the subscripts respectively correspond to the cases where $\delta=+\tfrac{\pi}{2}$ and $\delta=-\tfrac{\pi}{2}$.

It is evident from Eq. \er{eq:HOMIModes} that the field mode functions associated with photons exiting ports $c$ and $d$ are identical, and that they may be expressed in terms of the spin-orbit basis modes $\mathbfcal{E}_{00}$ defined in Eq. \er{eq:Amplitudes} as 

\beq{eq:PsiVector}
\bm{\Psi}_{\mp}\!\plr{r,\Phi}\eq\big(e_{00}\mathbfcal{E}_{00}+e_{11}\mathbfcal{E}_{11}\big)\mp i\big(e_{10}\mathbfcal{E}_{10}+e_{01}\mathbfcal{E}_{01}\big).
\eeq

 \begin{figure}
   \begin{center}
   \includegraphics[width=0.6\columnwidth]{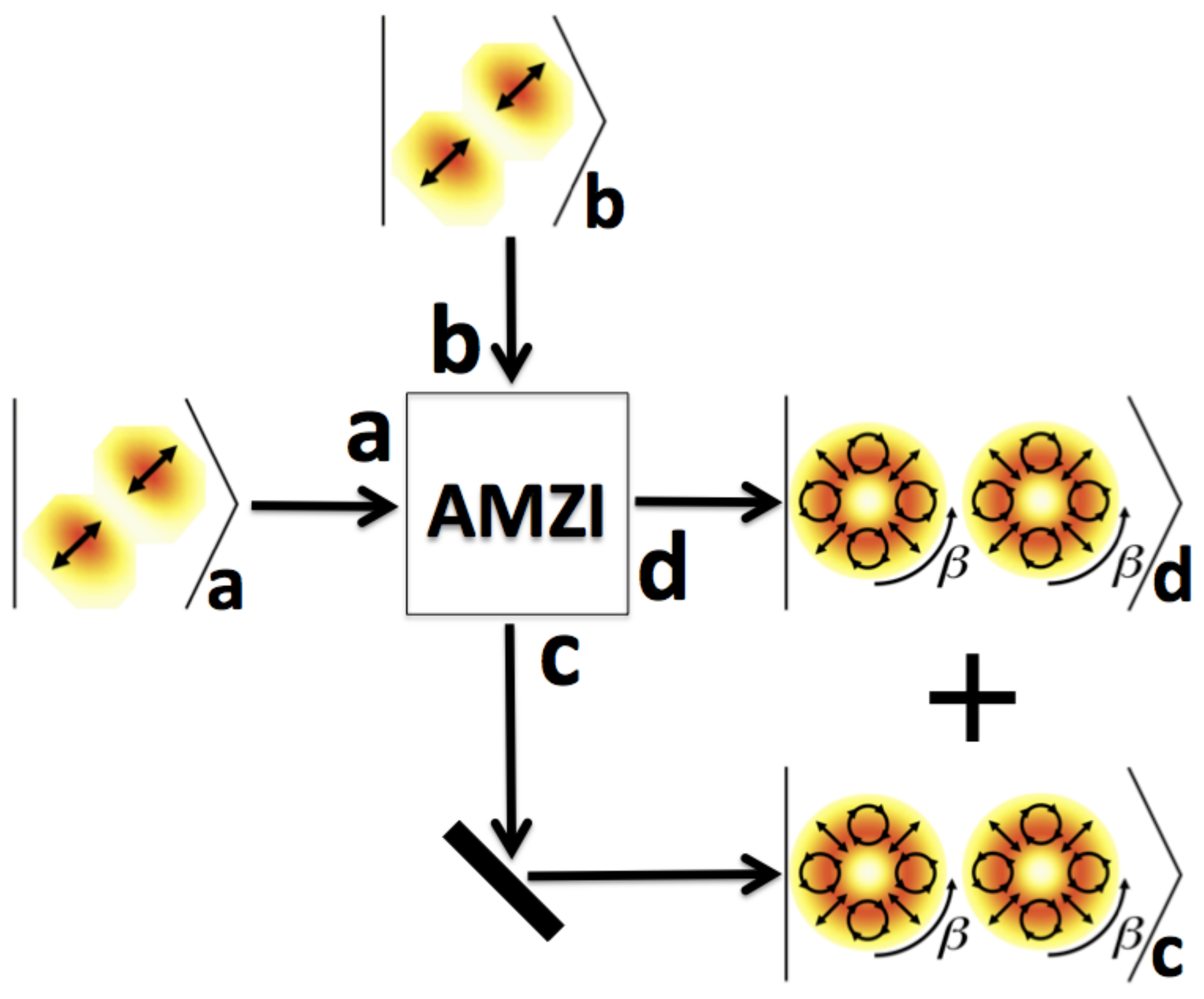} \\
   \end{center}
   \vspace{-3mm}
   \caption 
   {\label{fig:Biphoton_Interference} Correlated biphoton pairs prepared in indistinguishable spin-orbit product modes exhibit biphoton interference within the AMZI in conjunction with a vector mode conversion, with output Fock state $\ket{2_{c,\bm{\Psi}_{\mp}}0_{d,\bm{\Psi}_{\mp}}}-\ket{0_{c,\bm{\Psi}_{\mp}}2_{d,\bm{\Psi}_{\mp}}}$ (see Eq. \er{eq:BiphotonOutputState}).}  
   \end{figure} 
   
\no In contrast with the spin-orbit product modes given in Eq. \er{eq:Product}, the mode functions $\bm{\Psi}_{\mp}\!\plr{r,\Phi}$ are nonseparable, in the sense that they are not factorizable into products of their spin and orbital degrees of freedom.  Therefore, we conclude that our asymmetric Mach-Zehnder interferometer can exhibit Hong-Ou-Mandel type biphoton interference between input photons in indistinguishable spatially coherent spin-orbit product modes, while simultaneously implementing a mode conversion to vector modes of the form Eq. \er{eq:PsiVector}.  See Fig. \ref{fig:Biphoton_Interference} for a visualization of this where both input photons are in the mode shown in Fig. \ref{fig:Phenomenon}.

\section{Spin-orbit vector mode functions}\lb{sec:ModeFunctions}

\subsection{Classical interference of output fields} 

If the dual fields impinging on interferometer input ports $a$ and $b$ are chosen as mutually coherent laser beams as opposed to the correlated biphoton pairs discussed above, a classical description of the resulting interference between output fields is sufficient.  In this picture, dual input beams with spin-orbit product state fields of the form of Eq. \er{eq:Product} are converted to classically entangled vector outputs by the asymmetric Mach-Zehnder interferometer.  Assuming here that the input fields of Eq. \er{eq:Product} share a common overall classical amplitude $E_a=E_b\eq E$, we use Eqs. \er{eq:U} and \er{eq:EvenOdd} in order to calculate the output amplitudes $e_{jk}^{c}$ and $e_{jk}^{d}$.  Subsequently employing the spin-orbit product basis as in Eq. \er{eq:Inputs}, we calculate the output fields as

\begin{align} \lb{eq:DualInputOutputs}
\mathbfcal{E}_c\plr{r,\Phi}&= E\Big(\!\!\cos{\plr{\tfrac{\delta}{2}}}\!\!+\sin{\plr{\tfrac{\delta}{2}}}\!\Big)\bm{\Psi}_{-}\!\plr{r,\Phi}, \non \\
\mathbfcal{E}_d\plr{r,\Phi}&= E\Big(\!\!\cos{\plr{\tfrac{\delta}{2}}}\!\!-\sin{\plr{\tfrac{\delta}{2}}}\!\Big)\bm{\Psi}_{+}\!\plr{r,\Phi}.
\end{align}

\no where the mode functions $\bm{\Psi}_{+}$ and $\bm{\Psi}_{-}$ are given above in Eq. \er{eq:PsiVector}.  

From Eq. \er{eq:DualInputOutputs} it is evident that the output mode functions $\bm{\Psi}_{-}$ and $\bm{\Psi}_{+}$ remain stable at respective output ports $c$ and $d$ as the internal interferometer phase $\delta$ is tuned, while the overall output intensity at each port varies.  Furthermore, if the internal interferometer phase is adjusted such that $\delta=\pm\tfrac{\pi}{2}$---which yields Hong-Ou-Mandel type interference in the case of the aforementioned biphoton input---the classical description gives complete constructive interference at one output port and complete destructive interference at the other.  This stands in contrast to the quantum case of a biphoton input, where energy flows out of both outputs in the form of bunched photons according to Eq. \er{eq:BiphotonOutput}.

For an intuitive physical understanding of the conjunction of the phenomena of biphoton interference and mode conversion, it is instructive to consider the case of a single classical input beam directed into port $a$ only.  Taking this approach, we set the classical input field amplitudes in Eq. \er{eq:Product} according to $E_a=E$ and $E_b=0$, which implies that the $e_{jk}^{b}$ coefficients associated with port $b$ each must be zero in Eq. \er{eq:U}.  Substituting Eqs. \er{eq:EvenOdd} into Eq. \er{eq:U} with $e_{jk}^{b}=0$ and setting $\delta=\pm\tfrac{\pi}{2}$ as above then gives the relevant output amplitudes $e_{jk}^{c}$ and $e_{jk}^{d}$, which, when substituted into Eq. \er{eq:Inputs}, leads to the following expression for the output fields:

\bseq{eq:SingleInputOutputsHOMIConditionA}
\begin{align}
\mathbfcal{E}_c\plr{r,\Phi}=\pm\tfrac{1}{\sqrt{2}}\,\bm{\Psi}_{\mp}\!\plr{r,\Phi}, \\
\mathbfcal{E}_d\plr{r,\Phi}=+\tfrac{1}{\sqrt{2}}\,\bm{\Psi}_{\mp}\!\plr{r,\Phi},
\end{align}
\eseq

\no where the $+$ and $-$ signs correspond to $\delta=\pm\tfrac{\pi}{2}$, as before.  Repeating the same calculation for a single input directed into port $b$, we arrive at similar expressions:

\bseq{eq:SingleInputOutputsHOMIConditionB}
\begin{align}
\mathbfcal{E}_c\plr{r,\Phi}=+\tfrac{1}{\sqrt{2}}\,\bm{\Psi}_{\mp}\!\plr{r,\Phi}, \\
\mathbfcal{E}_d\plr{r,\Phi}=\mp\tfrac{1}{\sqrt{2}}\,\bm{\Psi}_{\mp}\!\plr{r,\Phi},
\end{align}
\eseq

\no which differ only with respect to the overall phase of the outputs.  Superposing the output fields given in Eqs. \er{eq:SingleInputOutputsHOMIConditionA} and \er{eq:SingleInputOutputsHOMIConditionB} then yields output fields identical to those of the dual input case given in Eq. \er{eq:DualInputOutputs} with the internal interferometer phase set to $\delta=\pm\tfrac{\pi}{2}$.  This result makes it evident that the constructive and destructive output interference that occurs in the dual input case is a consequence of interference \emph{between the two input beams}.  In contrast, the vector mode conversion effect occurs as a result of interference \emph{between the two interferometer paths} for each input beam separately, according to Eqs. \er{eq:SingleInputOutputsHOMIConditionA} and \er{eq:SingleInputOutputsHOMIConditionB}.  

In the quantum case of biphoton interference, a single spatially coherent photon entering a given port in a spin-orbit product mode ``interferes with itself'' at the output via the two interferometer paths, resulting in a conversion to the same vector mode functions $\bm{\Psi}_{\mp}$ as in the present classical case.  However, in the quantum optical picture, interference at a given output does not occur between \emph{between} single photons; interference occurs between quantum amplitudes associated with joint measurement outcomes that cannot distinguish the paths taken by the photons traversing the system. Therefore, dual input photons do not interfere at the outputs in the same way as dual classical input fields do in Eq. \er{eq:DualInputOutputs}.  Instead, the bunching phenomenon exhibited in Eq. \er{eq:BiphotonOutputState} results from destructive interference between the following joint outcomes, (i) the joint outcome in which the port $a$ and $b$ photons exit the respective ports $c$ and $d$ in field mode $\bm{\Psi}_{\mp}$, and (ii) the mutually indistinguishable joint outcome where these photons exit the opposite ports in the same field mode, which has an equal but opposite amplitude.

\subsection{Polarization and intensity distributions of spin-orbit vector output modes}\lb{sec:Vector}

We present below a theoretical and experimental analysis of the class of mode functions $\bm{\Psi}_{\mp}$ generated by the asymmetric Mach-Zehnder interferometer, which represent nonseparable spin-orbit vector modes of light.  By employing Eqs. \er{eq:Amplitudes}, the vector output vector modes $\bm{\Psi}_{\mp}\!\plr{r,\Phi}$ defined in Eq. \er{eq:PsiVector} may be expressed in terms of the spin and orbital Poincar\'e sphere angles $\alpha$, $\beta$, $\theta$, and $\phi$ of the input spin and orbital states as

\bseq{eq:Parameters}
\begin{align}
&\bm{\Psi}_{\mp}=\cos{\!\plr{\tfrac{\alpha}{2}}}\psi_{\mp}\,\bm{\hat{y}}\mp i\sin{\!\plr{\tfrac{\alpha}{2}}}e^{i\beta} \psi_{\pm}\,\bm{\hat{x}}, \lb{eq:ParametersA} \\
&=E\bklr{\cos{\!\plr{\tfrac{\alpha}{2}}}\plr{\sqrt{R_{\mp}^2+M_{\mp}^2}\;e^{i\bklr{\tan^{-1}\plr{\tfrac{M_{\mp}}{{R_{\mp}}}}}}}\bm{\hat{y}}
\mp i\sin{\!\plr{\tfrac{\alpha}{2}}}e^{i\beta}\plr{\sqrt{R_{\pm}^2+M_{\pm}^2}\;e^{i\bklr{\tan^{-1}\plr{\tfrac{M_{\pm}}{{R_{\pm}}}}}}}\bm{\hat{x}}}, \lb{eq:ParametersB} \\
&=E\,e^{i\bklr{\tan^{-1}\plr{\tfrac{M_{\mp}}{{R_{\mp}}}}}}\!\bklr{\cos{\!\plr{\tfrac{\alpha}{2}}}\;\sqrt{R_{\mp}^2+M_{\mp}^2}\bm{\hat{y}}
\mp i\sin{\!\plr{\tfrac{\alpha}{2}}}\sqrt{R_{\pm}^2+M_{\pm}^2}\;e^{i\bklr{\beta-\tan^{-1}\plr{\tfrac{R_{\pm}M_{\mp}-M_{\pm}R_{\mp}}{R_{\pm}R_{\mp}+M_{\pm}M_{\mp}}}}}\bm{\hat{x}}}, \lb{eq:ParametersC} \\
&=E G\!\plr{\!r\!}e^{\!i\bklr{\tan^{-1}\!\plr{\!\tfrac{M_{\mp}}{{R_{\mp}}}\!}\!}}\!\!\!\bklr{\!\cos{\!\plr{\tfrac{\alpha}{2}}}\!\sqrt{A_+\pm B\sin{\phi}}\;\bm{\hat{y}}
\mp i\sin{\!\plr{\tfrac{\alpha}{2}}}\!\sqrt{A_+\mp B\sin{\phi}}\;e^{i\bklr{\beta\mp\tan^{-1}\!\plr{\!\tfrac{B\cos{\phi}}{A_-}\!}\!}}\bm{\hat{x}}}\!\!. \lb{eq:ParametersD}
\end{align}
\eseq

\no In Eq. \er{eq:ParametersA}, 
$ \psi_-\eq E\,G\plr{r}\bklr{\cos{\!\plr{\frac{\theta}{2}}}\sin{\!\Phi}- i\sin{\!\plr{\frac{\theta}{2}}}\cos{\!\Phi}\,e^{i\phi}}$ and $ \psi_+\eq E\,G\plr{r}\left[\cos{\!\plr{\frac{\theta}{2}}}\sin{\!\Phi}\right.$ $\left.+i\sin{\!\plr{\frac{\theta}{2}}}\cos{\!\Phi}\,e^{i\phi}\right]$ characterize the transverse spatial dependence of $\bm{\Psi}_{\mp}$, while in Eq. \er{eq:ParametersB} the $ \psi$ functions have been expressed in complex polar form, with $R_-\eq\text{Re}\{ \psi_-\}$, $M_-\eq\text{Im}\{ \psi_-\}$, $R_+\eq\text{Re}\{ \psi_+\}$, and $M_+\eq\text{Im}\{ \psi_+\}$, respectively.  In Eq. \er{eq:ParametersC}, a phase has been factored out of $\bm{\Psi}_{\mp}$ as shown, and the identity $\tan^{-1}u-\tan^{-1}v=\tan^{-1}\plr{\frac{u-v}{1+uv}}$ has subsequently been used in the rightmost phase factor.  Substitution of the definitions of $R_-$, $M_-$, $R_+$, and $M_+$ into the term in square brackets then leads to Eq. \er{eq:ParametersD}, where 

\beq{eq:A&BSymbols}
A_{\pm}\eq\sin^2{\!\plr{\tfrac{\theta}{2}}}\cos^2{\!\Phi}\pm\cos^2{\!\plr{\tfrac{\theta}{2}}}\sin^2{\!\Phi}, \;\;\;\; \;\;\;\; \;\;\;\; \;\;\;\;  \;\;\;\; \;\;\;\;  B\eq\tfrac{1}{2}\sin{\theta}\sin{2\Phi}. 
\eeq

Equation \er{eq:ParametersD} is a central result of this work.  It presents the mode functions output by the asymmetric Mach-Zehnder interferometer as parameterized by the spin and orbital Poincar\'e sphere angles $\alpha$, $\beta$, $\theta$, and $\phi$ of the input spin and orbital qubits.  Since the radial dependence of the output mode functions is restricted to the function $G\plr{r}$, which factors out of the polarization state contained in square brackets, it is evident that the spatial polarization distribution of these vector modes possesses a radial symmetry.  However, the polarization state of the light generally exhibits spatial variation with azimuthal polar angle $\Phi$ through the $\Phi$ dependence of $A_{\pm}$ and $B$, which accounts for the classically entangled nature of these mode functions.  Furthermore, we note that the spatial distribution of polarization may be tuned by adjusting the polarization parameter $\beta$ of the input mode, which is easily achieved by tilting a piece of birefringent material.

As the intensity distribution of a given output mode is proportional to the classical cycle-averaged power density $\bar{P}_p\propto\mathbfcal{E}_p\plr{r,\Phi}\cdot\mathbfcal{E}_p^*\plr{r,\Phi}$ exiting output port $p$, we employ Eq. \er{eq:ParametersD} to calculate this quantity here for the single input case expressed in Eqs. \er{eq:SingleInputOutputsHOMIConditionA}-\er{eq:SingleInputOutputsHOMIConditionB}, with the internal interferometer phase set to $\delta=\pm\tfrac{\pi}{2}$:

\beq{eq:SingleInputOutputPower}
\bar{P}_c=\bar{P}_d\propto\tfrac{1}{2}E^2G\!\plr{r}^2\Big({A_{+}\pm B\sin{\phi}\cos{\alpha}}\Big).
\eeq

\no From the above relation it is evident that the transverse spatial intensity distribution associated with the orbital degree of freedom of the output modes is generally coupled to the polarization state of the input mode through the parameter $\alpha$.  In what follows we experimentally verify the aforementioned generation and tunability of the spatially varying polarization distribution, in addition to the polarization based control of the intensity distribution just discussed.

\section{Polarization based control of vector mode distributions: theory and experiment}\lb{sec:Experiment}

\subsection{Experimental apparatus}

   \begin{figure}[t!]
   \begin{center}
   \begin{tabular}{cc}
  \includegraphics[width=0.4\columnwidth]{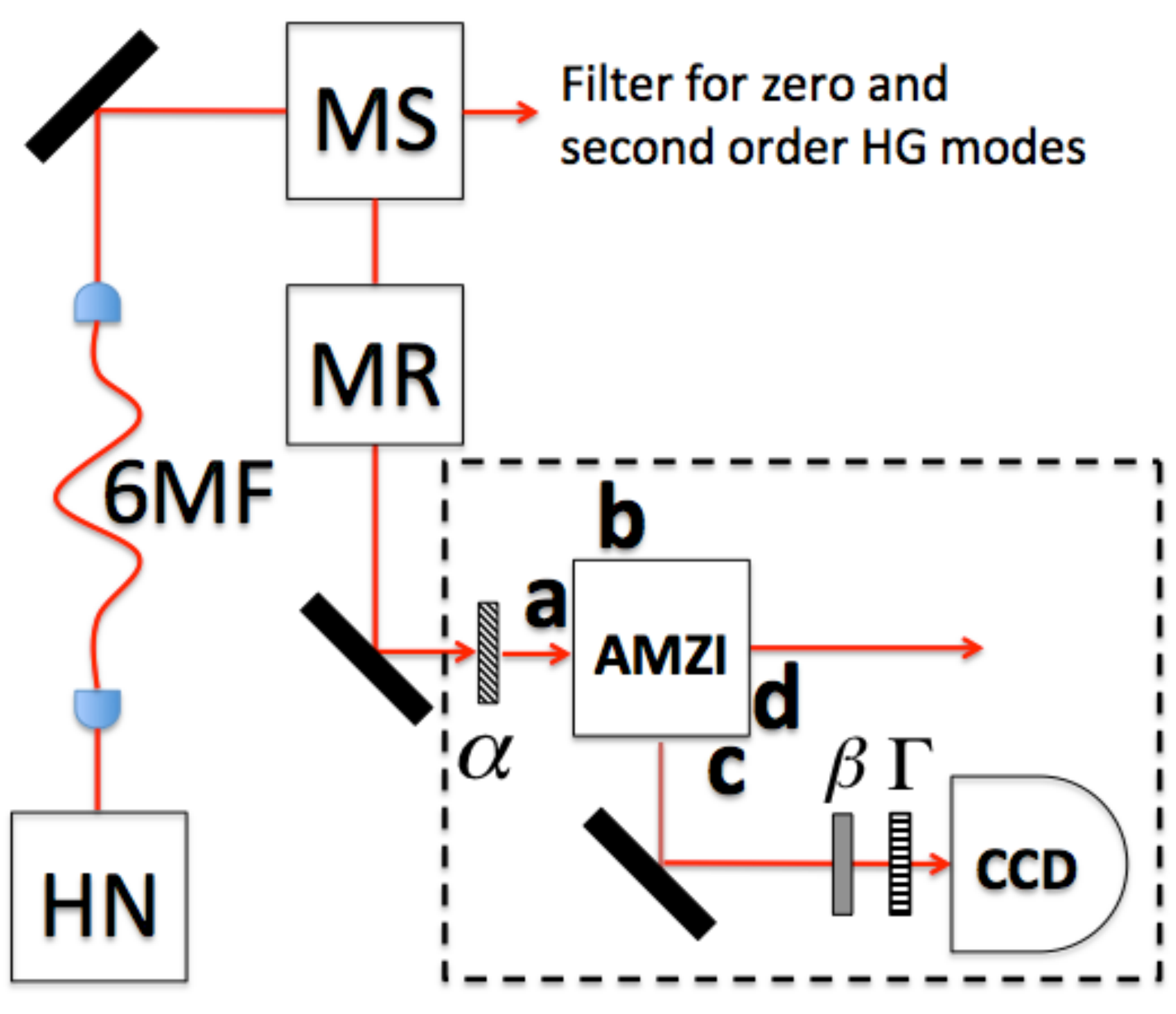} 
   &
  \includegraphics[width=0.4\columnwidth]{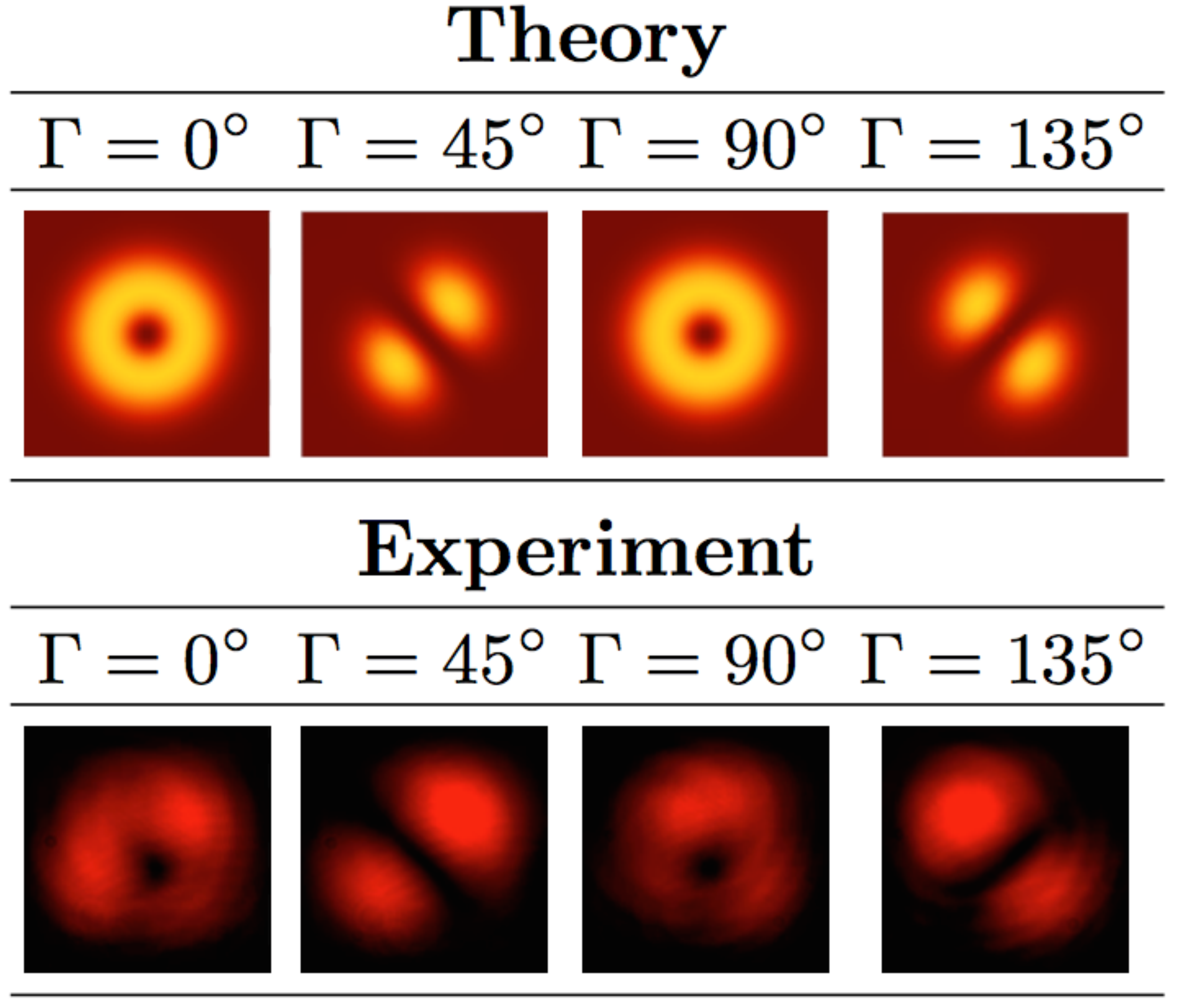}  \\
   \text{\Large{(\textbf{a})}} &    \text{\Large{(\textbf{b})}}  
   \end{tabular}
   \end{center}
   \vspace{-2.5mm}
   \caption 
   {\label{fig:Experiment1} 
(a) Experimental apparatus for conversion of spin-orbit product modes to vector modes.  The dashed lines surround the subset of the apparatus detailed in Fig. \ref{fig:Interferometer}.  Key: HN--helium neon Laser; 6MF--six mode fiber; MS--mode sorter; MR--mode rotator.  (b)  Linearly polarized intensity profiles of a spin-orbit vector output mode from a single spin-orbit product state input at interferometer port $a$, prepared in a balanced superposition states of both its spin and orbital degrees of freedom.  As the linear polarizer angle $\Gamma$ is rotated, the modal intensity distribution exhibits a  polarization-controlled oscillation between first order Laguerre-Gaussian and Hermite-Gaussian profiles, demonstrating the vector nature of the beam.
}
   \end{figure}

Our experimental apparatus is shown as Fig. \ref{fig:Experiment1}(a).  Light from a Helium Neon (HeNe) laser is coupled into an optical fiber that supports six transverse spatial modes for each polarization degree of freedom at the HeNe laser wavelength, all of which are generally excited with various amplitudes and relative phases.  The modes exiting the fiber are then filtered by means of a mode-sorting Sagnac interferometer which sorts the modes according to their two-dimensional parity (see \cite{Leary_2009} for details), such that the zero- and second-order fiber output modes are rejected, and only the two first-order modes remain in the beam.  The phase $\phi$ between the first-order transverse spatial modes is tuned as desired by imparting internal stress to the optical fiber via bending \cite{McGloin_1998}, while the the mode orientation variable $\theta$ is tuned via an optical mode rotator, which consists of three mirrors oriented to mimic the reflections experienced in a rotatable dove prism \cite{Galvez_1999}.  A linear polarizer or half wave plate near interferometer port $a$ then controls the polarization orientation angle $\alpha$ of the input beam, while a birefringent retarder see Fig. \ref{fig:Interferometer}) controls the polarization retardance $\beta$.  The balanced spin-orbit product mode inputs are then converted to vector output modes by the interferometer according to Eqs. \er{eq:SingleInputOutputsHOMIConditionA} and \er{eq:ParametersD}, after which their intensity distributions are measured at the outputs via CCD camera, while a linear polarizer is inserted in front of the CCD in order to analyze their polarization distributions.

\subsection{Tunable spatial polarization distribution}

We prepare a single input beam at port $a$ in a diagonal Hermite-Gaussian profile, such that $\theta=\tfrac{\pi}{2}$ and $\phi=\pi$ on the orbital Poincar\'e sphere.  We similarly prepare the spin degree of freedom in a balanced superposition of vertically and horizontally polarized states, such that $\alpha=\tfrac{\pi}{2}$, while leaving the relative phase $\beta$ between input polarization components as a adjustable parameter.  Under these conditions, the input mode takes the form shown in Fig. \ref{fig:Phenomenon}(a), such that Eqs. \er{eq:SingleInputOutputsHOMIConditionA}, \er{eq:ParametersD}, and \er{eq:A&BSymbols} yield the port $c$ output vector fields with the internal interferometer phase set to $\delta=\pm\tfrac{\pi}{2}$,

\beq{eq:BalancedOutput}
\pm\mathbfcal{E}_c\!\plr{r,\Phi}=\mathbfcal{E}_d\!\plr{r,\Phi}=\tfrac{E}{2\sqrt{2}}G\!\plr{r}e^{\mp i\Phi}\!\plr{\bm{\hat{y}}
\mp ie^{i\plr{\beta\pm2\Phi}}\bm{\hat{x}}}\!,
\eeq

\no while the output power density is given by $\bar{P}_{p}\plr{r}\propto\tfrac{1}{2}E^2G\!\plr{r}^2$-- ``donut'' modes for both output ports.  The vertical and horizontal components of the field in Eq. \er{eq:BalancedOutput} each bear unit orbital angular momentum, but with opposite signs, as may be seen by distributing the azimuthal phase term $e^{\mp i\Phi}$.  Due to the presence of the azimuthal polar coordinate $\Phi$ in the relative phase factor $e^{i\plr{\beta\pm2\Phi}}$, the output polarization distributions vary spatially with $\Phi$ as illustrated in Fig. \ref{fig:Phenomenon}(b).  Furthermore, the entire polarization pattern propagates about the phase singularity at the center of the spatial mode as $\beta$ is tuned, with the direction of rotation depending on the sign of $\delta$.

In order to analyze the spatially varying polarization distribution of the spin-orbit vector modes shown in Fig. \ref{fig:Phenomenon}(b), we insert linear polarizers at output ports $c$ and $d$, with their transmission axes aligned at an angle $\Gamma$ with respect to the vertical (as measured anticlockwise as viewed from the beam source).  The polarization components of the fields exiting the polarizer may then be found via the Jones calculus \cite{Jones_1941},

\begin{align} \lb{eq:Jones}
\left(\begin{array}{c} {\mathcal{E}_{p_{x}}\!\plr{\Gamma}}  \\ {\mathcal{E}_{p_{y}}\!\plr{\Gamma}} \end{array}\right)
&=\left(\begin{array}{cc} {\sin^2\Gamma} & {\sin{\Gamma}\cos{\Gamma}} \\ {\sin{\Gamma}\cos{\Gamma}} & {\cos^2\Gamma} \end{array}\right)
\left(\begin{array}{c} {\mathcal{E}_{p_{x}}}  \\ {\mathcal{E}_{p_{y}}} \end{array}\right),
\end{align}

\no where $\mathbfcal{E}_p\plr{r,\Phi}=N\plr{\mathcal{E}_{p_{x}}\bm{\hat{x}}+\mathcal{E}_{p_{y}}\bm{\hat{y}}}$, such that the polarization components $\mathcal{E}_{p_{x}}=\mp ie^{i\plr{\beta\pm2\Phi}}$ and $\mathcal{E}_{p_{y}}=1$, and common factor $N\eq\tfrac{E}{\sqrt{2}}G\!\plr{r}e^{\mp i\Phi}$, may be read off from Eq. \er{eq:BalancedOutput}.  Employing Eq. \er{eq:Jones}, we then calculate the power density of the polarized outputs for $\delta=\pm\tfrac{\pi}{2}$,

\begin{align}\lb{eq:PolarizedOutputs}
\bar{P}_p\plr{r,\Phi}&=\abs{N}^2\plr{\abs{\mathcal{E}_{p_{x}}\!\plr{\Gamma}}^2+\abs{\mathcal{E}_{p_{y}}\!\plr{\Gamma}}^2} \propto E^2 G\plr{r}^2\Big(1\pm\sin{\plr{2\Gamma}}\sin{\plr{\beta\pm2\Phi}}\Big).
\end{align}

   \begin{figure}[t!]
   \begin{center}
   \begin{tabular}{cc}
  \includegraphics[width=0.5\columnwidth]{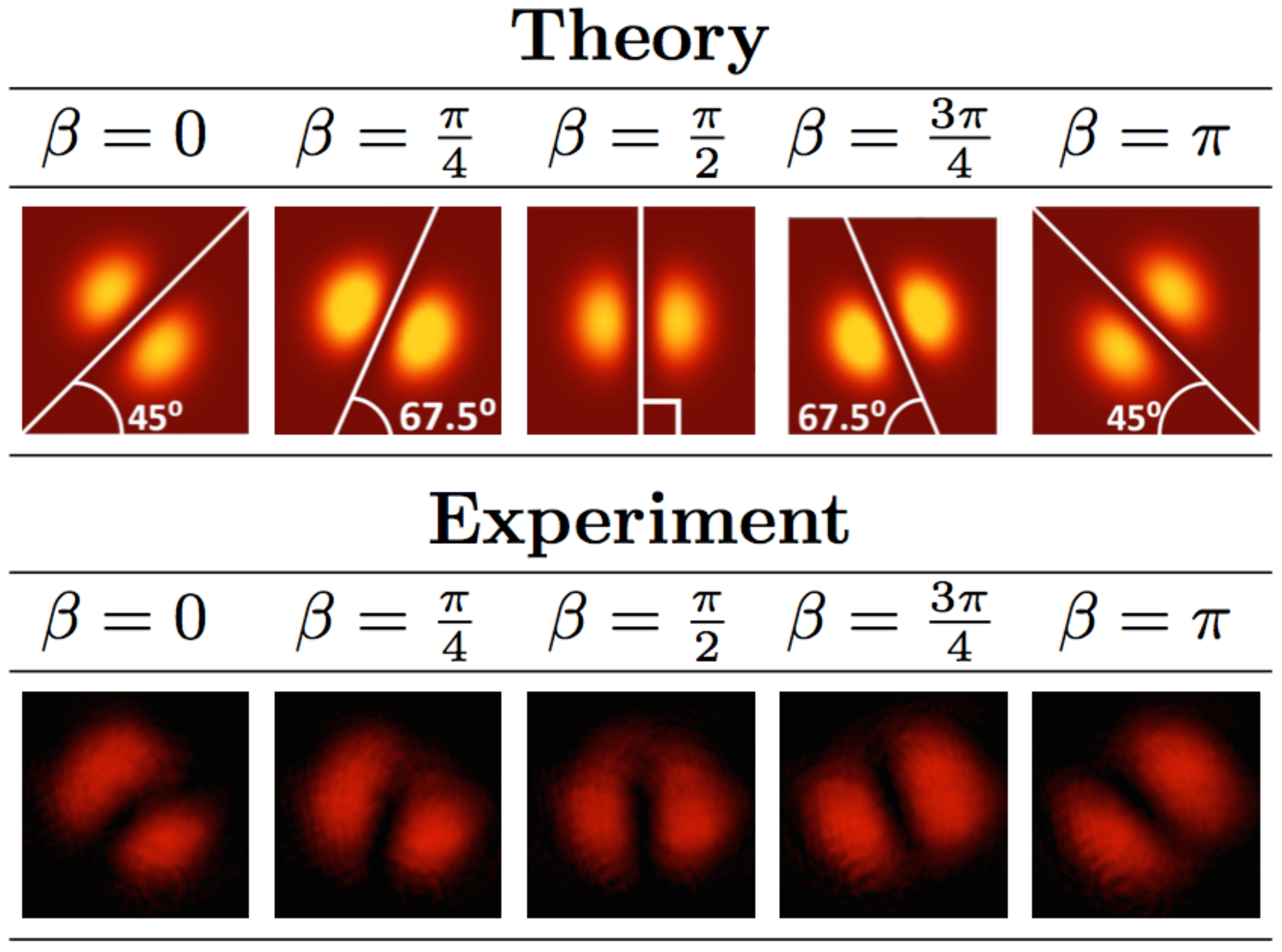} 
   &
  \includegraphics[width=0.315\columnwidth]{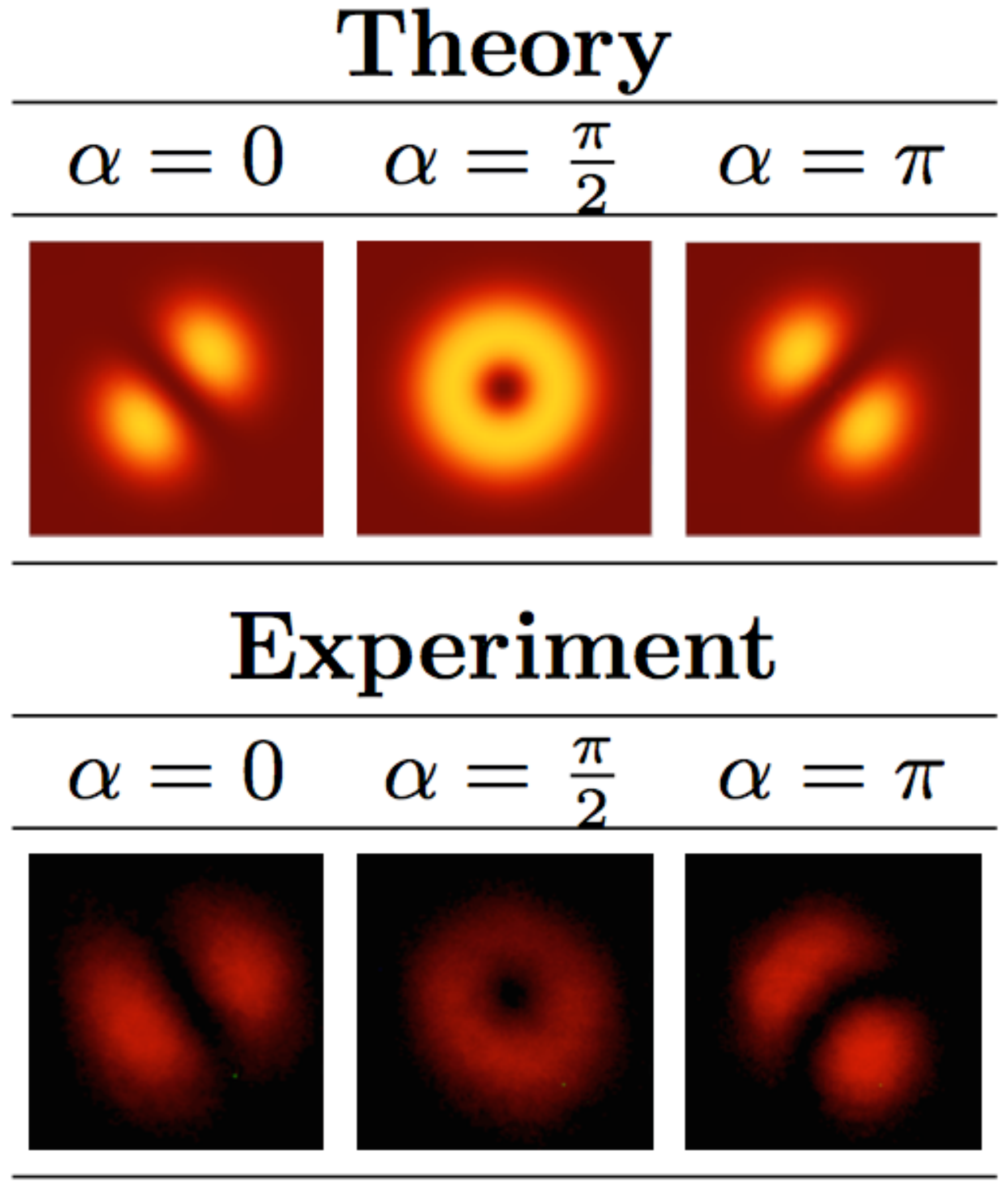}  \\
   \text{\Large{(\textbf{a})}} &    \text{\Large{(\textbf{b})}}  
   \end{tabular}
   \end{center}
   \vspace{-2.5mm}
   \caption 
   {\label{fig:Experiments2&3} 
{(a) Diagonally linearly polarized intensity profiles of a spin-orbit vector output mode from a similar input mode as in Fig. \ref{fig:Experiment1}(b) (see text for details).  As the retardance $\beta$ between horizontal and vertical polarization components is increased, the Hermite-Gaussian profile exhibits a polarization-controlled rotation.  (b) Intensity profiles of vector output modes with no linear polarizer present before the camera (see text for details), from a single spin-orbit product state input at interferometer port $a$, prepared in a balanced superposition state of its orbital degree of freedom.  As the polarization of the input mode is rotated from vertical to horizontal by means of a half wave plate, the modal intensity distribution exhibits a  polarization-controlled oscillation between Laguerre-Gaussian and Hermite-Gaussian profiles.}
}
   \end{figure}

To test the presence of the spatially varying polarization structure shown in Fig. \ref{fig:Phenomenon}(b) and reflected in Eq. \er{eq:PolarizedOutputs}, we vary the linear polarizer angle $\Gamma$ while holding $\beta$ fixed at $\pi$.  Theoretical density plots of the power density $\bar{P}_c\plr{r,\Phi}$ of the mode exiting port $c$ for various polarizer angles $\Gamma$ are included in Fig. \ref{fig:Experiment1}(b) and compared to our experimentally measured intensity distributions.  As can be seen from the figure, the modal intensity distribution oscillates between Laguerre-Gaussian and Hermite-Gaussian profiles as $\Gamma$ is varied.  This is consistent with Eq. \er{eq:PolarizedOutputs} with $\beta$ set to $\pi$, which predicts a radially symmetric ``donut'' distribution for $\Gamma=0^{\circ}$ and  $\Gamma=90^{\circ}$, while reducing to $\bar{P}_p\plr{r,\Phi}\propto 2E^2 G\plr{r}^2\cos^{2}\!\plr{\tfrac{\pi}{4}+\Phi}$ for $\Gamma=45^{\circ}$ and $\bar{P}_p\plr{r,\Phi}\propto 2E^2 G\plr{r}^2\sin^{2}\!\plr{\tfrac{\pi}{4}+\Phi}$ for $\Gamma=135^{\circ}$.  These are the diagonal and antidiagonal Hermite-Gaussian modes.   We conclude that this data supports the presence of the azimuthally varying polarization structure described in Eq. \er{eq:BalancedOutput}. 

In order to verify that the azimuthal propagation of this polarization structure is controlled by the polarization retardation parameter $\beta$ as reflected in  Eq. \er{eq:PolarizedOutputs}, we hold the linear polarizer angle $\Gamma$ fixed at $45^{\circ}$, and vary $\beta$ by tuning a Berek polarization compensator at output port $c$.  Under these conditions, Eq. \er{eq:PolarizedOutputs} reduces to $\bar{P}_p\plr{r,\Phi}\propto 2E^2 G\plr{r}^2\sin^{2}\!\plr{\tfrac{\pi}{4}+\tfrac{\beta}{2}+\Phi}$ for $\delta=\tfrac{\pi}{2}$, which predicts anticlockwise azimuthal rotation of the resulting linearly polarized Hermite-Gaussian spatial mode as $\beta$ is tuned.  This experiment is compared with theory in Fig. \ref{fig:Experiments2&3}(a), which includes output intensity distributions for port $c$ for various values of $\beta$.  Our observations of the patterns shown above serve as evidence in support of the presence of the vector output modes of the type described in Fig. \ref{fig:Phenomenon}, with the specific mode shown there corresponding to the $\beta=0$ case of Fig. \ref{fig:Experiments2&3}(a).

\subsection{Tunable spatial intensity distribution}

In the experiment of Fig. \ref{fig:Experiments2&3}(a), we tuned the polarization distribution of a spin-orbit vector mode while its intensity distribution (\emph{before} passing through the polarization analyzer) remained constant.  Here we produce a different type of vector mode whose intensity distribution varies as the input mode polarization orientation angle $\alpha$ is tuned.  

We prepare a single classical input beam at port $a$ in a balanced superposition state of its orbital degree of freedom only, such that $\theta=\tfrac{\pi}{2}$.  However, this time we stress the fiber to set $\phi=-\tfrac{\pi}{2}$, so that the transverse spatial distribution of the input field consists of a first-order Laguerre-Gaussian mode \cite{McGloin_1998}.  Under these conditions, Eqs. \er{eq:SingleInputOutputsHOMIConditionA}, \er{eq:ParametersD}, and \er{eq:A&BSymbols} yield the vector output vector fields with the internal interferometer phase again set to $\delta=\tfrac{\pi}{2}$,

\beq{eq:DonutInput}
\mathbfcal{E}_c\!\plr{r,\Phi}=\pm \tfrac{E}{\sqrt{2}}G\!\plr{r}\!\Big(\cos\!\plr{\tfrac{\alpha}{2}}\!H_{+} \, \bm{\hat{y}}-i\sin\!\plr{\tfrac{\alpha}{2}}\!H_{-} \, \bm{\hat{x}}\Big),
\eeq

\no where $H_{+}\plr{\Phi}\eq\cos\plr{\tfrac{\pi}{4}+\Phi}$ and $H_{-}\plr{\Phi}\eq\sin\plr{\tfrac{\pi}{4}+\Phi}$ respectively represent diagonal and antidiagonal first order Hermite-Gaussian modes.  The output power density associated with this mode may be calculated directly from the general Eq. \er{eq:SingleInputOutputPower} as

\beq{eq:VariableIntensity}
\bar{P}_{p}\plr{r,\Phi}\propto\tfrac{1}{4}E^2G\!\plr{r}^2\Big(1+\cos\plr{\alpha}\sin\plr{2\Phi}\Big),
\eeq

\no which predicts that the modal intensity distribution, being made up of orthogonal Hermite-Gaussian modes in orthogonal polarization states, may be tuned by varying $\alpha$.  Fig. \ref{fig:Experiments2&3}(b) compares this prediction with experiment for several values of $\alpha$, clearly demonstrating the coupling between the input polarization and the spatial intensity distribution of the output.

\section{Conclusions}

We have demonstrated experimentally the generation of vector modes of light via an asymmetric Mach-Zehnder interferometer with a spin-orbit product state input.  The modes generated comprise various special cases of Eq. \er{eq:ParametersD}, which parameterizes an entire class of nonseparable modes with radially symmetric polarization profiles in terms of the spin and orbital Poincare sphere angles of the input product states.  We have predicted that correlated input photon pairs in indistinguishable spin-orbit product states will exhibit biphoton (Hong-Ou-Mandel) interference in this apparatus, in conjunction with a mode conversion to vector modes of the form of Eq. \er{eq:ParametersD}, leading to path-entangled output photons in nonseparable spin-orbit vector modes.  

We note that cylindrically symmetric spin-orbit vector modes similar to those presented in Fig. \ref{fig:Experiment1} have been previously produced by others using a structured optical fiber \cite{Milione_2010}, and a number of interferometry-based schemes have been employed for the measurement and manipulation of similar mode types (e.g.\cite{Souza_2010,Vieira_2013,da_Silva_2016}.  In contrast with this previous work, here we have produced nonseparable modes in a four-port device capable of exhibiting biphoton interference in conjunction with a mode conversion from separable to nonseparable modes, demonstrated above in Figs. \ref{fig:Experiment1} and \ref{fig:Experiments2&3}.

Since the asymmetric interferometer exploits both second- and fourth-order coherence effects, it is capable of imposing transformations on the spin and orbital degrees of freedom of a spatially coherent input photon via single-photon interference, while simultaneously allowing for the occurrence of two-photon interference.  For this reason, standard biphoton sources, such as those from spontaneous parametric down conversion, must be appropriately spatially filtered in order to allow for the observation of such biphoton vector mode conversion effects.

We expect the tunable vector mode polarization and spatial field distributions predicted and demonstrated in this work to have a variety of applications to the field of light-matter interactions, ranging from the design of dynamic optical tweezers using focussed vector beams \cite{Skelton_2013}, to the impartation of information contained in structured light onto atomic systems \cite{Radwell_2015, Parigi_2015}, both at the single-photon level and for path-entangled biphoton states in structured vector modes.

\section*{Acknowledgments}

We acknowledge funding from the Research Corporation for Science Advancement by means of a Cottrell College Science Award, and the Henry Luce Foundation by means of the Clare Boothe Luce Program.

\end{document}